\documentclass[runningheads]{llncs}
\usepackage{amsmath}
\usepackage{graphicx}
\usepackage{tabularx}
\usepackage{booktabs}
\newtheorem{prop}{Proposition}
\usepackage{xcolor}
\usepackage{float}
\usepackage{amssymb}
\usepackage{appendix}
\usepackage{subfigure}
\usepackage{enumerate}

\begin{document}
\title{Differential Game Analysis for Cooperation Models in Automotive Supply Chain under Low-Carbon Emission Reduction Policies
\thanks{Yukun Cheng is the corresponding author. This work is supported by the National Nature Science Foundation of China (No. 11871366) and the Research Innovation Program for College Graduate Students of Jiangsu Province (No. KYCX22 3249).}}
\titlerunning{Cooperation Models in Automotive Supply Chain}
%
\author{Yukun Cheng\inst{1,3}\orcidID{0000-0002-3638-3440} \and Zhanghao Yao\inst{2}\and Xinxin Wang\inst{2}}
%
%
\institute{School of Business, Jiangnan University, Wuxi, 214122, China\\ \and School of Business, Suzhou University of Science and Technology,\\  \and Think Tank for Urban Development, Suzhou University of Science and Technology,\\ Suzhou, 215009, China\\ \email{ykcheng@amss.ac.cn, zhyao@post.usts.edu.cn, 2228052391@qq.com}}
\maketitle         
\begin{abstract}
In the context of reducing carbon emissions in the automotive supply chain, collaboration between vehicle manufacturers and retailers has proven to be an effective measure for enhancing carbon emission reduction within the enterprise. This study aims to evaluate the effectiveness of such collaboration by constructing a differential game model that incorporates carbon trading and consumer preferences for low-carbon products. The model examines the decision-making process of an automotive supply chain comprising a vehicle manufacturer and multiple retailers. By utilizing the Hamilton-Jacobi-Bellman equation, we analyze the equilibrium strategies of the participants under both a decentralized model and a Stackelberg leader-follower game model.
In the decentralized model, the vehicle manufacturer optimizes its carbon emission reduction efforts, while each retailer independently determines its low-carbon promotion efforts and vehicle retail price. In the Stackelberg leader-follower game model, the vehicle manufacturer cooperates with the retailers by offering them a subsidy. Consequently, the manufacturer plays as the leader, making decisions on carbon emission reduction efforts and the subsidy rate, while the retailers, as followers, compute their promotion efforts and retail prices accordingly.
Through theoretical analysis and numerical experiments considering the manufacturer's and retailers' efforts, the low-carbon reputation of vehicles, and the overall system profits under both models, we conclude that compared to the decentralized model, where each party pursues individual profits, the collaboration in the Stackelberg game yields greater benefits for both parties. Furthermore, this collaborative approach promotes the long-term development of the automotive supply chain.
\end{abstract}
%
%
%


\section{Introduction}

The automotive industry, in particular, has been identified as one of the major contributor to global carbon emissions, and as a result, has been the focus of regulatory and societal pressures to reduce its carbon emissions. In response to these pressures, many manufactures have set ambitious goals to reduce their carbon emissions, including improving energy efficiency and adopting electric vehicles \cite{al2021adoption}. However, achieving these goals is not always straightforward, as manufacturers operate in a complex ecosystem that includes not only themselves but also retailers, consumers, and other participants.

One important part in this ecosystem is the interaction between the manufacturers and their retailers.
The manufacturer is responsible for carbon emission reduction efforts by conducting research and development of low-carbon technologies and producing low-carbon vehicles. Retailers play a critical role in the distribution of vehicles, using low-carbon promotions to attract consumers to purchase low-carbon products. Besides, the manufacturer and retailers may also cooperate in their emission reduction activities. However, retailers and manufacturer are primarily driven by their own interests to maximize profits, which may lead to suboptimal resource allocation and market efficiency. Therefore, designing appropriate cooperation models between manufacturers and retailers is crucial. Generally, decentralized decision-making and the Stackelberg leader-follower game model are widely adopted \cite{wang2021carbon,sun2022differential,xu2023cost}.
In decentralized decision-making, manufacturers and retailers can make independent decisions based on their own information and considerations. This decision design encourages them to make optimal decisions based on their own interests. At the same time, decentralized decision-making can reduce the costs of information exchange and coordination.
Through the Stackelberg leader-follower decision-making, manufacturer can first formulate the carbon emission reduction effort and subsidy rate, while retailers react based on the manufacturer's decisions. This arrangement of decision sequences can fully leverage the manufacturer's advantage and achieve market efficiency to some extent.

When it comes to government efforts to reduce carbon emissions, various administrative measures are implemented, including carbon taxes, building carbon trading markets, increasing green investments, and more. Among these measures, carbon trading is the most widely used, which has been implemented in many countries, including China and the European Union \cite{xia2018carbon}. Under a carbon trading system, the government initially allocates a certain amount of carbon quotas to each enterprise based on specific allocation rules. If an enterprise's actual carbon emissions are lower than the initial quota, they can sell the excess quotas in the carbon trading market to generate profit. On the other hand, enterprises that exceed their carbon quotas need to buy additional quotas to comply with government regulations \cite{zang2020does}. It is evident that the carbon trading policy influences the production and operation decisions of enterprises \cite{benjaafar2012carbon}. Taking the impact of carbon trading into consideration, Yang et al. \cite{yang2017pricing} designed pricing and emission reduction models for two competitive supply chains under a carbon trading scheme. Sun et al. \cite{sun2022differential} studied multi-period continuous production subject to dynamically changing characteristic conditions within the framework of carbon trading, and proposed different differential game models to explore cooperation models between the government and enterprises. Unlike existing works that typically involve two participants, such as one manufacturer and one supplier or one manufacturer and one retailer, this paper focuses on studying the interactions among one manufacturer and multiple retailers.

Many studies have investigated the influence of consumers' low-carbon preferences on the supply chain. Xia et al. \cite{xia2018carbon}  incorporated reciprocal preferences and consumers' low-carbon awareness into a supply chain model consisting of a single manufacturer and a single retailer, and examined how these preferences affect the decisions, performance, and efficiency of the supply chain members. Wang et al. \cite{wang2021carbon} studied the carbon reduction decisions of automotive supply chain members under total control and transaction rules using differential game theory, where consumers' low-carbon preferences were also a significant factor affecting companies' emission reduction decisions. Chen et al. \cite{chen2023coordination} developed a closed-loop supply chain model for recycling and remanufacturing based on Stackelberg leader-follower game theory, considering consumers' low-carbon preferences and government subsidies, and established profit models for each stakeholder under centralized and decentralized decision-making models. They also examined the impact of consumers' low-carbon preferences on the profits and decisions of supply chain members. Xu et al. \cite{xu2023cost} developed a cost-sharing model for the automotive supply chain, considering the dynamic changes in consumers' low-carbon preferences. They argued that consumer demand would be affected by the low-carbon reputation of products, which in turn would impact product sales. Similar insights can also be found in the studies of \cite{sun2022differential,bian2020manufacturer,zhang2021dynamic}, which suggest that market demand depends on a company's green reputation and the level of environmental friendliness of its products, and that the formation of a green reputation requires coordination among supply chain members. These studies highlight that consumers' low-carbon preferences have become an essential factor in supply chain decision-making. Additionally, we noted that \cite{wang2021carbon,xu2023cost} considered dynamic consumer low-carbon preferences in their research. Dynamic preferences are more reflective of their impact on the supply chain compared to static preferences. Therefore, drawing inspiration from \cite{sun2022differential,xu2023cost}, our paper aims to introduce dynamic consumer low-carbon preferences based on green low-carbon reputation.

In reality, the process of reducing carbon emissions within the supply chain is a dynamic and long-term endeavor. Throughout this process, the level of emission reduction and the green low-carbon reputation of vehicles, as well as consumers' low-carbon preferences, are continuously changing. This prompts us to adopt a dynamic perspective to study the interaction between the manufacturer and $n$ retailers. Differential game theory, as an important dynamic game model, can provide solutions for dynamic equilibrium outcomes in continuous time \cite{li2020research}. By utilizing the Hamilton-Jacobi-Bellman equation, we analyze the equilibrium strategies of participants in two distinct models: a decentralized model and a Stackelberg leader-follower game model.
In the decentralized model, each participant makes individual decisions, whereas in the Stackelberg game model, the manufacturer engages in collaboration with the retailers by offering a subsidy. Unlike previous studies that mainly focus on the manufacturer's strategy of carbon emission reduction efforts and the retailers' strategies for low-carbon promotion, our study goes further by incorporating the retailers' decision-making process regarding retail pricing.
Furthermore, to enhance the realism of our model, we introduce carbon trading as an additional element.

The remaining parts of this article are organized as follows. Section 2 provides the introduction of  the problem and the necessary assumptions. In Section 3, two differential cooperation models are established and the corresponding equilibrium solutions are analyzed. Section 4 conducts the experiments and the numerical analysis to verify our theoretical results. Finally, Section 5 presents the conclusion and discussion on further directions. The detailed proofs of the main results are placed in the full version.

\section{Problem assumptions and notations}

This paper focuses on a two-tier automotive supply chain consisting of a manufacturer and $n$ retailers.
The manufacturer is responsible for carbon emission reduction efforts to develop low-carbon technology and producing vehicles under a brand, and its investment in low-carbon technology can enhance the low-carbon reputation of its products and the emissions reduction level.
The retailers strive to improve the low-carbon reputation of the brand of vehicles through carbon promotion efforts, such as publicity, subsidies, and other efforts to attract more consumers. Both the manufacturer and the retailers' efforts can improve the low-carbon reputation of this brand of vehicle. In addition, due to the existence of cap-and-trade regulation, the manufacturer needs to consider the impact of carbon trading when making decisions.

Considering the interaction between the manufacturer and the retailers, we propose two cooperation models: the decentralized model and the Stackelberg leader-follower game model, to explore the strategy selection and interaction among the participants, respectively. Each model accounts for the interdependence between the manufacturer and the retailers, as well as the impact of carbon trading and consumer green preferences on their decision-making process.

Tabel \ref{tab1} provides the notations used in this paper.
\begin{table}[!ht]\label{tab1}
  \centering
  \caption{Notations and description.}
   \begin{tabularx}{340pt}{cX}
\toprule
             \textbf{Notation}                   &    \textbf{Descriptions}
\\ \hline
     $t$       &  Time period
\\
       $G(t),G(0)$                  &     Low carbon reputation of the vehicle at time $t$,  and initial value of the low carbon reputation, $G(0) \geq 0$.
\\
        $E_m, E_{r_i}$          & Manufacturer's carbon emission reduction effort, retailer $i$'s low-carbon promotion effort.
\\
        $\lambda_m, \lambda_{r_i}$          & Manufacturer's cost coefficient related to carbon emission reduction, retailer cost coefficient related to the promotion of low-carbon vehicle, $\lambda_m, \lambda_{r_i}> 0$.
\\
      $\mu_m, \mu_{r_i}$              &    Influence coefficient of manufacturer emission reduction efforts on the reputation, influence coefficient of retailer's low-carbon promotion efforts on reputation, $\mu_m, \mu_{r_i}>0$.
\\
     $\omega$
                    &   Influence coefficient of manufacturer emission reduction efforts on the emission reduction level, $\omega>0$.
\\
       $p,\ p_i,\ p_c$          &  Manufacturer's wholesale price, retail price, price per unit of carbon emission credit, $p, p_i, p_c>0$.
\\
          $Q_i(t)$               &    Demand function for retailer $i$ at time $t$ and the total demand is $Q(t)=\sum_{i=1}^nQ_i(t)$.
\\
     $\theta$         &   Low-carbon preference of consumers, $\theta>0$.
\\
      $ a_i $              &  retailer $i$'s potential sales, $ a_i >0$.
\\
       $ b_i,c$          & Influence coefficient of price on sales, influence coefficient of other retailers' prices on sales, $ b_i>0,0\leq c\leq 1$.
\\
     $F_0$         &   Carbon emission quota.
\\
      $ F(t) $              & Total quantity of carbon quota trading at time $t$.
\\

       $ \delta$          &   Decay coefficient of low-carbon reputation, $ \delta>0$.
\\
     $\rho$         &    Discount rate of profit, $\rho>0$ .
\\
\bottomrule
\end{tabularx}
\end{table}

\subsection{Model Assumptions}

\paragraph{\bfseries Assumption 1.}
Both the carbon emission reduction effort $E_m$ of the manufacturer and the low-carbon promotion effort $E_{r_i}$ of retailer $i$ affect the low-carbon reputation $G$ of the brand of vehicles. Similar to \cite{xu2023cost}, this dynamic process of $G(t)$ is described by the following differential equation:
{\small
\begin{equation*}
\begin{aligned}
\dot{G}(t)=\mu_m E_m(t)+\sum_{i=1}^n \mu_{r_i} E_{r_i}(t)-\delta G(t),
\end{aligned}
\end{equation*}}
where $\mu_m$ and $\mu_{r_i}$, $i=1,\cdots,n$, are  the influence coefficients of manufacturer's and retailers' efforts on the reputation, $\delta>0$ is the decay coefficient of low-carbon reputation.

\paragraph{\bfseries Assumption 2.}
By assumptions for demand in \cite{zhang2014strategic}, the demand $Q_i$ of the vehicles sold by retailer $i$ is decreasing with the retail price $p_i$ and increasing with the price $p_j$, $j\neq i$ set by others. In addition, the higher lower-carbon reputation $G(t)$ can positively influence demand $Q_i(t)$. Thus
{\small
\begin{equation*}
\begin{aligned}
Q_i(t)=(a_i-b_i p_i+c \sum_{k=1,k\neq i}^{n} \frac{b_k p_k}{n-1}) \theta G(t),
\end{aligned}
\end{equation*}}
where $b_i>0$ is the coefficient of the effect from the retail price $p_i$ on demand $Q_i$. Since the effect from other retail price $p_k\neq p_i$ on $Q_i$ is no more than the effect from $p_i$, we let the coefficient $0\leq c\leq 1$.

\paragraph{\bfseries Assumption 3.}
The carbon emissions level of the vehicle is determined by the manufacturer's efforts in emission reduction (through low-carbon technology investment and development), with the emission reduction level being proportional to the manufacturer's efforts, denoted by a coefficient $\omega$. Thus, the vehicle's emission reduction level is equal to $\omega \cdot E_m(t)$. Similar to the assumption in \cite{xia2018carbon,wang2021carbon}, we assume that the initial carbon emissions per unit product is 1, indicating the emission reduction per unit product is $1 \cdot \omega \cdot E_m(t)$, the per unit product carbon emission is $1 - \omega \cdot E_m(t)$, and the total emission of the manufacturer is $(1-\omega E_m(t))\sum_{i=1}^nQ_i(t)$. Let us denote the carbon quota of the manufacturer at time $t$ as $F(t)$, with $F_0$ being the initial carbon quota. Therefore, the carbon quota of the manufacturer is expressed as:
{\small
\begin{equation*}
\begin{aligned}
F(t)=F_0-(1-\omega E_m(t)) \sum_{i=1}^n Q_i(t).
\end{aligned}
\end{equation*}}
It is worth to note that when $F(t) > 0$, the manufacturer possesses excess carbon quotas that can be traded for financial gain in the carbon market. Conversely, if $F(t) < 0$, the manufacturer is obligated to procure carbon quotas from the market to comply with the government's regulations.

\paragraph{\bfseries Assumption 4.}

The costs of emission reduction efforts paid by the manufacturer and retailer are quadratic functions of $E_m(t)$ and $E_{r_i}(t)$, respectively, which are popular and have been adopted by a lot of literatures \cite{wang2021carbon,sun2022differential,xia2018carbon}. Therefore, the costs of efforts respectively paid by the manufacture and retailer $i$ are
{\small
\begin{equation*}
C_m(t)=\frac{1}{2} \lambda_m E_m^2(t),\ \  C_{r_i}(t)=\frac{1}{2} \lambda_{r_i} E_{r_i}^2(t).
\end{equation*}}

\paragraph{\bfseries Assumption 5.}
Due to the fact that the $n$ retailers are the retailers of the same brand of automobile vehicles, which are provided by the manufacturer, their business scale, target customers, and sales models are all similar. Therefore, we suppose these $n$ retailers to be homogeneous, and thus the cost coefficients, the impact coefficients of effort levels on reputation, and the price sensitivity coefficient are all set to be the same, that is,  $\lambda_r = \lambda_{r_1} = \lambda_{r_2} = \ldots = \lambda_{r_i}$, $\mu_r = \mu_{r_1} = \mu_{r_2} = \ldots = \mu_{r_i}$, $b = b_1 = b_2 = \ldots = b_i$.


\section{Model formulation}

Based on the assumptions in Section 2, this section establishes two differential game models to analyze the equilibrium strategies of the manufacture and the retailers, by considering the long-term impact of carbon trading and low-carbon reputation of vehicles.
For the sake of convenience, we omit $t$ in the following.

\subsection{Decentralized model}
Before establishing the decentralized model, it is necessary for us to clarify that the objectives of the manufacture and the retailers are all to maximize their own profits individually over an infinite time, and the discount rate is denoted by $\rho > 0$. Therefore, their objective functions are formulated as follows.
{\small\begin{eqnarray}
\max J_m^D & =&\int_0^{\infty} e^{-\rho t} \left[p  \sum_{i=1}^n Q_i  + p_c \left(F_0-(1-\omega E_m\right) \sum_{i=1}^n Q_i)
-\frac{1}{2} \lambda_m E_m^2 \right]dt, \label{jmD}  \\
\max J_{r_i}^D & =&\int_0^{\infty} e^{-\rho t}\left[p_i Q_i -\frac{1}{2} \lambda_r E_{r_i}^2\right]dt,\label{jrD}
\end{eqnarray}}
where $p$ is the wholesale price, $p_c$ is the price of per carbon credit, which are given in advance; and $p_i$ is the retail price set by retailer $i$.
Let $V_m^D$, $V_{r_i}^D$ denote the profit function of manufacture and retailer $i$, respectively. Thus, the corresponding  Hamiltonian-Jacobi- Bellman (HJB) equations are formulated as:
{\small
\begin{eqnarray}
\rho V_m^D &=& \max \bigg[p \sum_{i=1}^n Q_i + p_c (F_0-(1-\omega E_m) \sum_{i=1}^n Q_i) - \frac{1}{2} \lambda_m E_m^2+V_m^{D^{\prime}}(\mu_m E_m \nonumber \\
&+&\sum_{i=1}^n \mu_r E_{r_i}-\delta G)\bigg], \label{vmd} \\
\rho V_{r_i}^D &=&\max \bigg[p_i Q_i -\frac{1}{2} \lambda_r E_{r_i}^2 + V_{r_i}^{D^{\prime}} (\mu_m E_m+\sum_{i=1}^n \mu_r E_{r_i}-\delta G) \bigg]. \label{vrd}
\end{eqnarray}}
Let us denote $q_i=a_i-b p_i + c \sum_{k=1,k \neq i}^{n} \frac{b p_k}{n-1}$, meaning that $Q_i(t)=\theta G(t) q_i$. So
{\small
\begin{eqnarray}
\sum_{i=1}^n q_i=\sum_{i=1}^n (a_i-b p_i + c \sum_{k=1,k \neq i}^{n} \frac{b p_k}{n-1})=\sum_{i=1}^na_i-(1-c)b\sum_{i=1}^np_i.
\end{eqnarray}}

\begin{prop}\label{prop1} Under the decentralized model, the optimal carbon emission reduction effort of the manufacturer and the optimal low-carbon promotion efforts of retailers are
{\small
\begin{eqnarray}
{E_m^D}^* =\frac{p_c  \theta \omega  G  \sum_{i=1}^n q_i + (2A^D G+B^D)\mu_m}{\lambda_m},\ \
{E_{r_i}^D}^* =\frac{\mu_r D_i^D}{\lambda r},
\end{eqnarray}}
the optimal retailer price set by retailer $i$ is
{\small
\begin{eqnarray}
{p_i^D}^* =\frac{(n-1)(2-c) a_i+c \sum_{k=1}^n a_{k}}{b(2-c)(2 n-2+c)},\label{pid*}
\end{eqnarray}}
the optimal trajectory of low carbon reputation is:
{\small
\begin{eqnarray}
G^{D*} =\frac{  4 B^D \lambda_r \mu_m^2 + 4 \lambda_m \mu_r^2 \sum_{i=1}^n D_i^D}{ \lambda_r \sqrt{ \triangle^D }- \lambda_m \lambda_r \rho },
\end{eqnarray}}
and the optimal profits of manufacture and retailer $i$ are:
{\small
\begin{eqnarray}\label{as}
     V_m^D= A (G^{D*})^2 + B G^{D*} + C^D,\ \ V_{r_i}^D= D_i^D G^{D*} + H_i^D,
\end{eqnarray}}
where $A^D, B^D, C^D$ and $D_i^D$ are the coefficients in the value functions $V_m^D(G)= A^D G^2 + B^D G + C^D $ and $V_{r_i}^D= D_i^D G + H_i^D$,
{\small
\begin{eqnarray}
A^D &=&\frac{ 4 \lambda_m \delta + 2 \lambda_m \rho-4 p_c \theta \omega \mu_m \sum_{i=1}^n q_i - \sqrt{ \triangle^D }}{8 \mu_m^2},\label{ad}\\
B^D &=& \frac{4 \lambda_m  \lambda_r \theta (p-p_c) \sum_{i=1}^n q_i +  8 A \lambda_m \mu_r^2\sum_{i=1}^n  D_i }{ 2 \lambda_m \lambda_r \rho + \sqrt{ \triangle^D }}, \label{bd}\\
C^D&=& \frac{p_c F_0}{\rho} +\frac{{B^D}^2 \mu_m^2}{2 \lambda_m \rho}+ \frac{B^D \sum_{i=1}^n \mu_r E_{r_i}}{\rho},\label{cd}\\
D_i^D &=&\frac{\lambda_m p_i q_i \theta}{\lambda_m(\rho +\delta) - p_c \theta \omega \mu_m \sum_{i=1}^n q_i -2 A \mu_m^2},\label{dd}
\end{eqnarray}}
where $\triangle^D = (4 p_c \theta \omega \mu_m \sum_{i=1}^n q_i-4 \lambda_m \delta-2\lambda_m \rho)^2 - 16(\mu_m p_c \theta \omega  \sum_{i=1}^n q_i)^2 > 0$.
\end{prop}
The proof of Proposition \ref{prop1} is provided in the \ref{appendix1}.


\subsection{Stackelberg leader-follower game model}
In the Stackelberg leader-follower game model, the manufacturer supports the retailers by offering a subsidy. In this model, the manufacturer  plays as a leader to disclose its carbon emission reduction effort $E_m$ and subsidy rate $x_i$ to retailer $i$ in the first stage, and its objective function is
{\small
\begin{eqnarray}
\max J_m^S&=&\int_0^{\infty} e^{-\rho t} \bigg[p  \sum_{i=1}^n Q_i  + p_c (F_0-(1-\omega E_m) \sum_{i=1}^n Q_i) -\frac{1}{2} \lambda_m E_m^2\nonumber\\
&-&\frac{1}{2} \sum_{i=1}^n \lambda_r x_i E_{r_i}^2 \bigg]dt.
\end{eqnarray}}
Then, in the second stage retailer $i$ determines its low-carbon promotion effort $E_{r_i}$ after observing the manufacturer actions as a follower. Therefore, the objective functions of retailer $i$ is given by:
{\small
\begin{eqnarray}\label{jms}
\max J_{r_i}^S=\int_0^{\infty} e^{-\rho t}\left[p_i Q_i -\frac{1-x_i}{2} \lambda_r E_{r_i}^2\right]dt.
\end{eqnarray}}
Let $V_m^S$, $V_{r_i}^S$ denote the value functions of manufacture and retailer $i$, respectively. We have the Hamiltonian-Jacobi- Bellman (HJB) equations as
{\small
\begin{eqnarray}
\rho V_m^S&=&\max \bigg[p \sum_{i=1}^n Q_i + p_c (F_0-(1-\omega E_m) \sum_{i=1}^n Q_i) - \frac{1}{2} \lambda_m E_m^2 -\frac{1}{2} \sum_{i=1}^n \lambda_r x_i E_{r_i}^2 \nonumber\\
&+& V_m^{S^{\prime}}(\mu_m E_m+\sum_{i=1}^n \mu_r E_{r_i}-\delta G) \bigg], \label{vms} \\
\rho V_{r_i}^S&=&\max \bigg[p_i Q_i -\frac{1-x_i}{2} \lambda_r E_{r_i}^2 + V_{r_i}^{S^{\prime}} (\mu_m E_m+\sum_{i=1}^n \mu_r E_{r_i}-\delta G)\bigg].\label{vrs}
\end{eqnarray}}

\begin{prop}\label{prop2} Under the Stackelberg leader-follower game model, the equilibrium carbon emission reduction effort of
the manufacturer and the low-carbon promotion effort of retailer $i$ are presented as follows:
{\small
\begin{eqnarray}\label{ems*}
{E_m^S}^*=\frac{p_c  \theta \omega G  \sum_{i=1}^n q_i + (2A^S G+B^S)\mu_m}{\lambda_m},\ \ {E_{r_i}^S}^*&=\frac{\mu_r D_i^S}{\lambda_r (1-x_i)},
\end{eqnarray}}
the equilibrium retailer price $p^{S*}_i$ is
{\small
\begin{eqnarray}
{p_i^S}^*=\frac{(n-1)(2-c) a_i+c \sum_{k=1}^n a_{k}}{b(2-c)(2 n-2+c)},
\end{eqnarray}}
the equilibrium subsidy rate $x^{*}_i$ is
{\small
\begin{eqnarray}
x_i^*=\frac{2 (A^S G_s^S + B^S) - D_i^S}{2 (A^S G_s^S + B^S)+ D_i^S},
\end{eqnarray}}
the equilibrium trajectory of low carbon reputation is:
{\small
\begin{equation}\label{G}
    \begin{aligned}
    G^{S*}=\frac{ 4 B^S \lambda_r \mu_m^2 + 4 n B^S\lambda_m \mu_r^2 + 2 \lambda_m \mu_r^2 \sum_{i=1}^n D_i^S  }{ \sqrt{ \triangle^S }- \lambda_m \lambda_r \rho}.
    \end{aligned}
\end{equation}}
and the equilibrium profits of manufacture and retailer $i$ under the Stackelberg leader-follower game are
{\small
\begin{eqnarray}
     V_m^S= A^S {G_s^D}^2 + B^S G_s^S + C^S,\  \ V_{r_i}^S= D_i^S G_s^S + H_i^S,
\end{eqnarray}}
whereh $A^S, B^S, C^S$ and $D_i^S$ are the coefficients of the value functions $V_m^S= A^S G^2 + B^S G + C^S$ and $ V_{r_i}^S= D_i^S G + H_i^S$, with the formulations as:
{\small
\begin{eqnarray}
     A^S&=&\frac{(4 \delta \lambda_m \lambda_r  + 2 \lambda_m \lambda_r \rho -4   \lambda_r \mu_m  p_c  \omega \theta \sum_{i=1}^n q_i ) - \sqrt{ \triangle^S }}{ 8 ( \lambda_r \mu_m^2  + n\lambda_m \mu_r^2 )},\\
     B^S &=& \frac{ 4 \lambda_m  \lambda_r  \theta  (p-p_c)   \sum_{i=1}^n q_i+ 4 A \lambda_m \mu_r^2 \sum_{i=1}^n D_i  }{  2 \lambda_m \lambda_r \rho +  \sqrt{ \triangle^S }},\\
     C^S&= &\frac{p_c F_0}{\rho} +\frac{{B^S}^2 \mu_m^2}{2 \lambda_m \rho} +\frac{ 4 n B^2 \mu_r^2 - 4 B^S \mu_r^2 \sum_{i=1}^n {D_i^S}^2 + \mu_r^2 \sum_{i=1}^n D_i^S}{8  \lambda_r \rho },\\
     D_i^S&=&\frac{ 4 \lambda_m \lambda_r \theta p_i q_i}{2 \lambda_m \lambda_r \rho + A^S \lambda_m \mu_r^2 + \sqrt{ \triangle^S }},
\end{eqnarray}}
where
{\small $$\triangle^S = (4  \lambda_m \lambda_r \delta + 2 \lambda_m \lambda_r \rho -4   \lambda_r \mu_m  p_c  \omega \theta \sum_{i=1}^n q_i )^2 - 16 \lambda_r (  \lambda_r \mu_m^2  + n \lambda_m \mu_r^2 ) ( p_c \theta \omega \sum_{i=1}^n q_i)^2.$$}
\end{prop}

The proof of Proposition \ref{prop2} is provided in \ref{appendix2}.

\section{Numerical analysis}
This section performs numerical experiments to assess the models' validity, conduct the sensitivity analysis of key parameters, and provide managerial insights. Moreover, we simulate the impact of changes in low-carbon reputation and supply chain members' profits under the scenario without carbon trading, aiming to examine the decision-makings under various policy conditions.

Drawing upon the works of Wang et al. \cite{wang2021carbon} and Xu et al. \cite{xu2023cost}, we set the relevant parameters for our numerical experiments as:
$G(0)=0,
\lambda_m=500,
\lambda_r=100,
\mu_m=2,
\mu_r=0.5,
\omega=0.4,
p=15,
\theta=0.6,
b=0.9,
c=0.8,
F_0=500,
\delta=0.8,
\rho=0.6,
n=6.$

We distinguish the existence of carbon trading policy by setting $p_c=1>0$ and the scenario without carbon trading by setting $p_c=0$.

The results of the identical decision-making model are depicted using the same color scheme. Specifically, the decentralized model is represented by the color cyan, while the Stackelberg game model is represented by the color blue. Furthermore, solid and dashed lines are employed to differentiate between cases with and without carbon trading, respectively. Finally,, we use the superscript $'N'$ to denote the scenario where the carbon trading policy is not implemented.

\subsection{Changes in manufacturer's and retailer's profits over time}

This section discusses the variation of profits for the manufacturer and the retailer over time for the decentralized model and the Stackelberg game model respectively in Fig. 1-(a) and Fig. 1-(b).


From the figures, we can observe that conditional cooperation leads to improved profits for both manufacturers and retailers. This is achieved through manufacturers providing subsidies to retailers to incentivize the adoption of more low-carbon promotional measures. Additionally, in the absence of carbon trading, both decision-making models result in lower profitability compared to the scenarios with carbon trading. This indicates that manufacturers derive benefits from engaging in carbon trading.

It is worth noting that in the decentralized model with carbon trading, retailers experience higher profitability than in scenarios without carbon trading. This suggests that retailers benefit from the enhanced reputation resulting from manufacturers' carbon emission reduction efforts. Overall, if the vehicles produced by the manufacturer are environmentally friendly, they can generate more profits from the carbon trading market and allocate subsidies to support retailers' promotion efforts. By channeling a portion of the carbon trading revenue into the supply chain, both the manufacturer and retailers are incentivized to actively participate in low-carbon management, ultimately leading to the decarbonization of the entire supply chain.

\begin{figure}[H]
  \centering
  \label{manufacture}
  \subfigure[Manufacturer's profits.]{\includegraphics[width=0.49\textwidth]{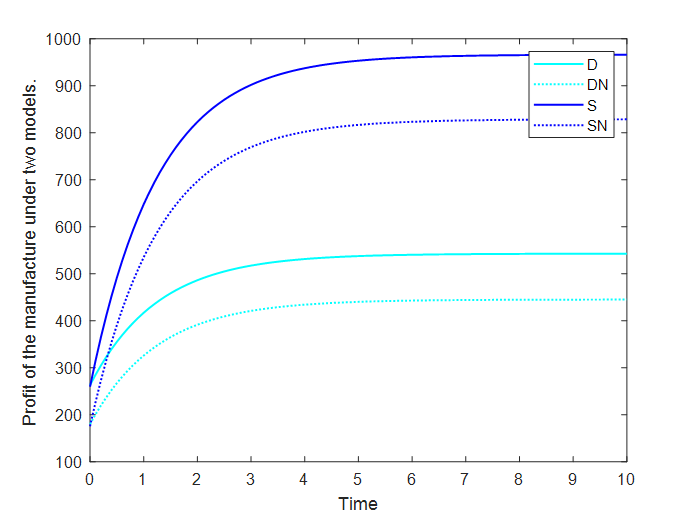}}
  \hfill
  \label{retailer}
  \subfigure[Retailers' profits.]{\includegraphics[width=0.49\textwidth]{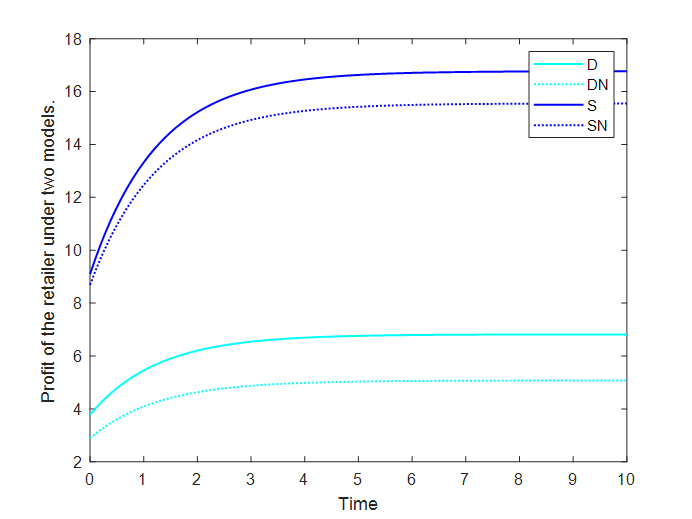}}
  \caption{Manufacturer's and retailers' profits under different models. }
\end{figure}

\subsection{Changes of low-carbon reputations and supply chain profits over time}

\begin{figure}[H]
  \centering
  \label{manufacture}
  \subfigure[Low-carbon reputation under different models.]{\includegraphics[width=0.49\textwidth]{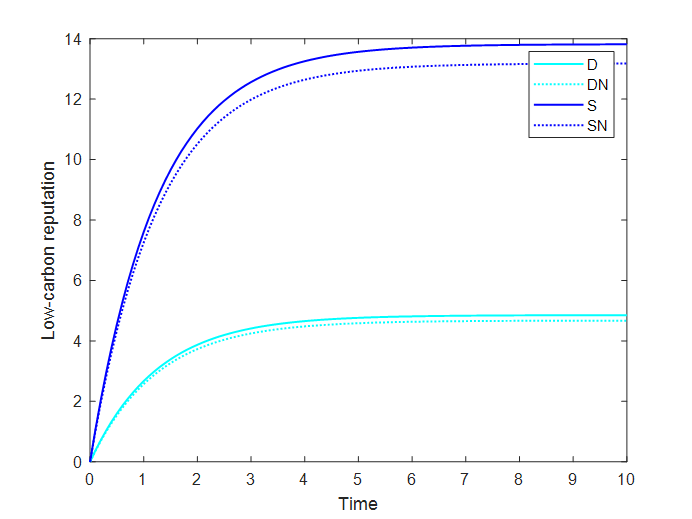}}
  \hfill
  \subfigure[Total profits of supply chain under different models.]{\includegraphics[width=0.49\textwidth]{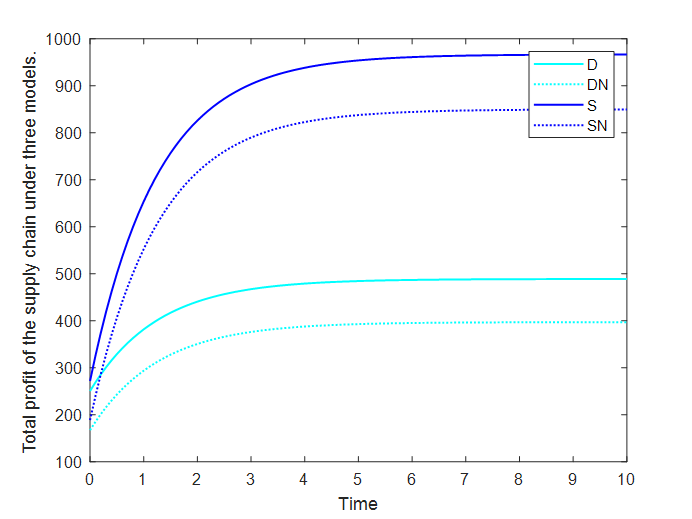}}
  \caption{Low-carbon reputations and total profits of supply chain under different models.}
\end{figure}

Fig. 2 illustrates the trajectories of low-carbon reputation under three scenarios, which increase with the increase of time $t$ and stabilize as $t$ approaches infinity.
In Fig. 2-(a), we can observe  that the low-carbon reputation under the Stackelberg game model is higher than that under the decentralized model,  which indicating that cooperation can enhance low-carbon reputation even if it is one-way. Furthermore, reputations were improved across all models where carbon trading was implemented. Therefore, supply chains should collaborate rather than act independently in the search for low-carbon solutions and the government should consider carbon trading as an alternative policy after subsidy cancellation.

From the Fig. 2-(b), it can be seen that the profit of supply chain in the Stackelberg game model is higher than that in the decentralized model, which demonstrates a similar trend as the low-carbon reputation shown in Fig. 2-(a).
These findings highlight the importance of prioritizing cooperation among supply chain members in the long run, as it can yield better profit performance and cost reduction. Moreover, the implementation of carbon trading enhances supply chain profits. Therefore, it is imperative for supply chain members to remain vigilant and adapt to policy changes accordingly.


\section{Conclusion}
This study employs a differential game framework to examine the dynamic decision-making process related to low-carbon strategies within the automotive supply chain. Departing from previous assumptions of fixed marginal profit, this study places emphasis on the joint decision-making regarding pricing and low-carbon operations. By considering the influence of carbon trading policies and the reputation of low-carbon vehicles, two interaction models between the manufacturer and $n$ retailers are discussed: the decentralized model and the Stackelberg game model. For each model, the equilibrium decisions of all participants and the trajectory of low-carbon reputation are analyzed. The key conclusions derived from this study are as follows:
\begin{itemize}
    \item  In the Stackelberg game model, where the manufacturer plays as the leader, providing subsidies to retailers to incentivize their low-carbon promotion efforts, several positive outcomes are observed. Compared to decentralized decision-making, the Stackelberg game model leads to improvements in supply chain profit, effort levels, and low-carbon reputation. This demonstrates that conditional cooperation through the Stackelberg model can effectively motivate supply chain members to actively engage in emission reduction activities.
    \item The low-carbon reputation of vehicles has a significant impact on market demand. Both retailers and manufacturers can increase their respective efforts to enhance the reputation of their vehicles, expand market share, and boost revenue.
    \item The implementation of carbon trading policies contributes to carbon emission reduction within the automotive supply chain and enhances the industry's low-carbon reputation to a certain extent. However, it is important to note that the initial adoption of carbon trading may impose challenges on the supply chain. Therefore, it is crucial for the government to dynamically adjust relevant policies to ensure a smooth transition towards a low-carbon supply chain.
\end{itemize}

However, this study does have some limitations that should be acknowledged. Firstly, the model settings do not incorporate the participation of retailers in carbon trading, which could be a valuable aspect to explore in future research. Secondly, the decision-making process within the automotive supply chain can be influenced by various other government policies, such as subsidies and carbon taxes, which were not extensively considered in this study. Finally, the research primarily focuses on manufacturers' direct retail channels and does not delve into the analysis of dual-channel sales, which could provide additional insights and avenues for further investigation.


%
%

\bibliographystyle{unsrt}
\bibliography{reference}

\appendix

\renewcommand{\thesection}{Appendix \Alph{section}}

\section{Proof for the decentralized model}\label{appendix1}

From the second derivative, we know that the right hand side of Eqs. (\ref{vmd}) is a concave function about $E_m$. Let the first derivative equal to zero, the manufacture's response function can be obtained as:

{\small
\begin{align}\label{emd}
E_m^D=\frac{p_c \omega \theta G  \sum_{i=1}^n q_i  +V_m^{D^{\prime} }\mu_m}{\lambda_m}.
\end{align}}

Then, we know that the right hand side of Eqs. (\ref{vrd}) is a concave function about $E_m, p_i$, respectively. Let the first derivative equal to zero, the retailer's response function can be obtained as:
{\small
\begin{align}
E_{r_i}^D&=\frac{\mu_r V_{r_i}^{D^{\prime}}}{\lambda r},\label{erd}\\
p_i^D&=\frac{a_i+c \sum_{i=1, i\neq n}^n \frac{b p_k}{n-1}}{2 b}.\label{pid}
\end{align}}

From Eqs. (\ref{pid}) and $0<c<1, n \geq 1$ we can deduce that:
\small{
\begin{equation}\label{pid1}
\begin{aligned}
2b \sum_{i=1}^n p_i^{D^*}=\sum_{i=1}^n a_i+c  b \sum_{k=1}^n p_k^{D^*} \Rightarrow \sum_{i=1}^n p_i^{D^*}=\frac{\sum_{i=1}^n a_i}{b(2-c)} \Rightarrow \sum_{k=1, k \neq i}^n p_k^{D^*}=\frac{\sum_{i=1}^n a_i}{b(2-c)}-p_i^{D^*}.
\end{aligned}
\end{equation}}

Substituting Eqs. (\ref{pid1}) into Eqs. (\ref{pid}), we can get:
\begin{equation}\label{pid2}
\begin{aligned}
p_i^D=\frac{(n-1)(2-c) a_i+c \sum_{k=1}^n a_k}{b(2-c)(2 n-2+c)}.
\end{aligned}
\end{equation}

Substituting Eqs. (\ref{emd}) and (\ref{erd}) into Eqs. (\ref{vmd}) and (\ref{vrd}), we can obtain:

\small{
\begin{equation}\label{vmd1}
\begin{aligned}
\rho V_m^D &= p \theta G \sum_{i=1}^n q_i  + p_c \bigg(F_0-(1-\omega E_m) \sum_{i=1}^n q_i \theta G \bigg) -\frac{1}{2} \lambda_m \bigg(\frac{p_c \omega \theta G  \sum_{i=1}^n q_i + V_m^{D^{\prime}}\mu_m}{\lambda_m} \bigg)^2 \\
& +V_m^{D^{\prime}} \bigg(\mu_m \frac{p_c \omega \theta G  \sum_{i=1}^n q_i + V_m^{D^{\prime}}\mu_m}{\lambda_m} + \sum_{i=1}^n \frac{\mu_r^2 V_{r_i}^{D^{\prime}}}{\lambda_r}-\delta G \bigg),
\end{aligned}
\end{equation}

\begin{equation}\label{vrd1}
\rho V_{r_i}^D= p_i q_i \theta G -{\frac{\mu_r^2 V_{r_i}^{D^{\prime}}}{2 \lambda_r}}^2 + V_{r_i}^{D^{\prime}} \bigg( \frac{p_c \mu_m \omega \theta G \sum_{i=1}^n q_i +V_m^{D^{\prime} }\mu_m}{\lambda_m}+\sum_{i=1}^n \frac{\mu_r^2 V_{r_i}^{D^{\prime}}}{\lambda_r}-\delta G \bigg).
\end{equation}

}

According to Eqs. (\ref{vmd1}) and (\ref{vrd1}), we further infer that the solution of the manufacturer's HJB equation is quadratic about $G$ and the solution of the retailer's HJB equation is linear about $G$. Therefore, we set $V_m^D= A^D G^2 + B^D G + C^D , V_{r_i}^D= D_i^D G + H_i^D$, and $V_m^{D^{\prime}}=2A^DG+B^D, V_{r_i}^{D^{\prime}}=D_i^D$ and substitute them into Eqs. (\ref{vmd1}) and (\ref{vrd1}). Subsequently, we obtain:

\small{
\begin{equation}\label{vmd2}
\begin{aligned}
&\rho (A^D G^2 + B^D G + C^D)=  \Bigg[\biggl( \frac{\left(p_c \theta w \sum_{i=1}^n q_i+2 A \mu m\right)^2}{2 \lambda_m}-2 A \delta \biggl)G^2 \\
& + \left(\left(p-p_c\right) \theta \sum_{i=1}^n q_i+\frac{B p_c \theta w \mu_m \sum_{i=1}^n q_i + 2 A B \mu_m^2}{\lambda_m}  + 2 A \sum_{i=1}^n \mu_r E_{r i}- B \delta \right)G  \\
&  + p_c F_0 +\frac{B^2 \mu_m^2}{2 \lambda_m}+B \sum_{i=1}^n \mu_r E_{r i} \Bigg],
\end{aligned}
\end{equation}

\begin{equation}\label{vrd2}
\begin{aligned}
 \rho ( D_i^D G + H_i^D)&=  (p q_i \theta +D_i^D ( \frac{p_c \omega \theta G \mu_m  \sum_{i=1}^n q_i + 2A^D G \mu_m^2}{\lambda_m} -\delta) )G  -{\frac{\mu_r^2 {D_i^D}^2}{2 \lambda r}}  \\
& + \frac{B^D D_i^D \mu_m}{\lambda_m} +\sum_{i=1}^n \frac{\mu_r^2 D_i^2}{\lambda r}.
\end{aligned}
\end{equation}}

Comparing the quadratic coefficient of $G$ in Eqs. (\ref{vmd2}), we obtain a quadratic equation with respect to $A^D$. Solving this equation yields the value of $A^D$ as shown in Eqs. (\ref{ad}). Noted that when $A^D=\frac{ 2 \lambda_m \delta + \lambda m p-4 p_c \theta \omega \mu_m \sum_{i=1}^n q_i + \sqrt{ \triangle^D }}{8 \mu_m^2} $, both parties' efforts are less than zero. However, in reality, the effort level cannot be negative. Therefore, $A^D$ equals to $\frac{ 2 \lambda_m \delta + \lambda m p-4 p_c \theta \omega \mu_m \sum_{i=1}^n q_i - \sqrt{ \triangle^D }}{8 \mu_m^2} $. Furthermore, by comparing the coefficients of other terms, we obtained the values of $B^D$, $C^D$, $D_i^D$ and $H_i^D$. are determined by the following equation:

{\small
\begin{align}
A^D &=\frac{ 4 \lambda_m \delta + 2 \lambda_m \rho-4 p_c \theta \omega \mu_m \sum_{i=1}^n q_i - \sqrt{ \triangle^D }}{8 \mu_m^2},\label{ad}\\
B^D &= \frac{4 \lambda_m  \lambda_r \theta (p-p_c) \sum_{i=1}^n q_i +  8 A \lambda_m \mu_r^2\sum_{i=1}^n  D_i }{ 2 \lambda_m \lambda_r \rho + \sqrt{ \triangle^D }}, \label{bd}\\
C^D&= \frac{p_c F_0}{\rho} +\frac{{B^D}^2 \mu_m^2}{2 \lambda_m \rho}+ \frac{B^D \sum_{i=1}^n \mu_r E_{r_i}}{\rho},\label{cd}\\
D_i^D &=\frac{\lambda_m p_i q i \theta}{\lambda_m(\rho +\delta) - p_c \theta \omega \mu_m \sum_{i=1}^n q_i -2 A \mu_m^2},\label{dd} \\
H_i^D&=-{\frac{\mu_r^2 {D_i^D}^2}{2 \lambda_r \rho}} +\frac{B^D D_i^D \mu_m^2}{\lambda_m \rho} + \frac{\mu_r^2  D_i \sum_{i=1}^n D_i}{2 \lambda_r \rho}.\label{ed}
\end{align}}

where $\triangle^D = (4 p_c \theta \omega \mu_m \sum_{i=1}^n q_i-4 \lambda_m \delta-2\lambda_m \rho)^2 - 16(\mu_m p_c \theta \omega  \sum_{i=1}^n q_i)^2 > 0$.

Substituting the optimal efforts of manufacture and retailer above into the dynamic function $\dot{G}(t)=\mu_m E_m+\sum_{i=1}^n \mu_r E_{r_i}-\delta G(t)$, we can obtain the steady-state low carbon reputation and the trajectory of low carbon reputation over time as follows:

\small{
\begin{equation}
    \begin{aligned}
    G_s^D=\frac{\mu_m E_m^D+\sum_{i=1}^n \mu_r E_{r_i}^D}{\delta}=\frac{  4 B^D \lambda_r \mu_m^2 + 4 \lambda_m \mu_r^2 \sum_{i=1}^n D_i^D}{ \lambda_r \sqrt{ \triangle^D }- \lambda_m \lambda_r \rho },
    \end{aligned}
\end{equation}

\begin{equation}
    \begin{aligned}
    G(t)^D = G_s^D+(G(0)-G_s^D) e^{-\delta t}.
    \end{aligned}
\end{equation}
}

Besides, based on the study of \cite{wang2021carbon}, the following conditions need to be satisfied in order to ensure $V_m^D(G)= A G^2 + B G + C$ and $V_{r_i}^D= D_i G_s^D + H_i$ equal to the value function:

\begin{equation}
    \begin{aligned}
        \lim _{t \rightarrow \infty} V_m^D(G) e^{-\delta t} =0.
    \end{aligned}
\end{equation}

Finally, by substituting $G_s^D$ into $V_m^D(G)$, we obtain the net profit of manufacture and retailer:

{\small
\begin{align}
V_m^D&= A {G_s^D}^2 + B G_s^D + C,\\
V_{r_i}^D&= D_i G_s^D + H_i.
\end{align}}








This completes the proof.
$\blacksquare$

\section{Proof of Stackelberg Game Model}\label{appendix2}

According to the backwards induction method, we first take the derivative of Eqs. (\ref{vrs}) with respect of $E_{r_i}$ and $p_i$. Let the first derivative equal to zero, the  response function can be obtained as:
{\small
\begin{align}
E_{r_i}^S&=\frac{\mu_r V_{r_i}^{S^{\prime}}}{\lambda_r (1-x_i)}, \label{ers} \\
p_i^S&=\frac{a_i+c \sum_{i=1, i\neq n}^n \frac{b p_k}{n-1}}{2 b}. \label{pis}
\end{align}}
From Eqs. (\ref{pis}) and $0<c<1, n \geq 1$ we can deduce that:
\small{
\begin{equation}\label{pid1s}
\begin{aligned}
2b \sum_{i=1}^n p_i^{S^*}=\sum_{i=1}^n a_i+c  b \sum_{k=1}^n p_k^{S^*} \Rightarrow \sum_{i=1}^n p_i^{S^*}=\frac{\sum_{i=1}^n a_i}{b(2-c)} \Rightarrow \sum_{k=1, k \neq i}^n p_k^{S^*}=\frac{\sum_{i=1}^n a_i}{b(2-c)}-p_i^{S^*}.
\end{aligned}
\end{equation}}
Substituting Eqs. (\ref{pid1s}) into Eqs. (\ref{pis}), we can derive the following equation:
\small{
\begin{equation}
    {p_i^S}=\frac{(n-1)(2-c) a_i+c \sum_{k=1}^n a_{k}}{b(2-c)(2 n-2+c)}.
\end{equation}}
Then, we substitute Eqs. (\ref{ers}) and Eqs. (\ref{pis})  into Eqs. (\ref{vms})  and take the derivative of $E_m$ and $x_i$ equal to zero.
{\small
\begin{align}
E_m^S&=\frac{p_c \omega \theta G  \sum_{i=1}^n q_i +V_m^{S^{\prime} }\mu_m}{\lambda_m}, \label{ems} \\
x_i&=\frac{2 V_m^{S^{\prime}} - V_{r_i}^{S^{\prime}}}{2 V_m^{S^{\prime}}  + V_{r_i}^{S^{\prime}}}. \label{xs}
\end{align}}
Substituting Eqs. (\ref{ers}) , (\ref{ems}) and (\ref{xs}) into Eqs. (\ref{vms}) and (\ref{vrs}), we can obtain:
\small{
\begin{equation}\label{vms1}
\begin{aligned}
\rho V_m^S=& p \theta G \sum_{i=1}^n q_i  + p_c (F_0-(1-\omega E_m) \sum_{i=1}^n Q_i)
- (\frac{(p_c \omega \theta G  \sum_{i=1}^n q_i + V_m^{S^{\prime}}\mu_m)^2}{2 \lambda_m})
-\frac{\mu_r^2 \sum_{i=1}^n (4 {V_m^{S^{\prime}}}^2 -  {V_{r_i}^{S^{\prime}}}^2)}{8 \lambda_r } \\
& +V_m^{S^{\prime}} ( \frac{p_c \mu_m \omega \theta G \sum_{i=1}^n q_i +V_m^{S^{\prime}}\mu_m^2}{\lambda_m} +  \frac{\mu_r^2  \sum_{i=1}^n (2 V_m^{S^{\prime}} -  V_{r_i}^{S^{\prime}})}{2 \lambda_r } -\delta G),
\end{aligned}
\end{equation}
\begin{equation}\label{vrs1}
\begin{aligned}
\rho V_{r_i}^S=& p_i \sum_{i=1}^n Q_i
-{\frac{\mu_r^2 V_{r_i}^{S^{\prime}} (2 V_m^{S^{\prime}} - V_{r_i}^{S^{\prime}} )}{4 \lambda_r }}\\
&+ V_{r_i}^{S^{\prime}}\left( \frac{p_c \mu_m \omega \theta G \sum_{i=1}^n q_i +V_m^{S^{\prime} }\mu_m}{\lambda_m}+ \frac{\mu_r^2 V_{r_i}^{S^{\prime}} \sum_{i=1}^n (2 V_m^{S^{\prime}} -  V_{r_i}^{S^{\prime}})}{2 \lambda_r }-\delta G\right).
\end{aligned}
\end{equation}
}
According to Eqs. (\ref{vms1}) and (\ref{vrs1}), we further infer that the solution of the manufacturer's HJB equation is quadratic about $G$ and the solution of the retailer's HJB equation is linear about $G$. Hence, we set $V_m^S= A^S G^2 + B^S G + C^S , V_{r_i}^S= D_i^S G + H_i^S$, and $V_m^{S^{\prime}}=2A^S G+B^S, V_{r_i}^{S^{\prime}}=D_i^S$, by comparing the coefficients on both sides of the equation, we get the following values:
{\small
\begin{align}\label{as}
     A^S&=\frac{(4 \delta \lambda_m \lambda_r  + 2 \lambda_m \lambda_r \rho -4   \lambda_r \mu_m  p_c  \omega \theta \sum_{i=1}^n q_i ) - \sqrt{ \triangle^S }}{ 8 ( \lambda_r \mu_m^2  + n\lambda_m \mu_r^2 )},\\
     B^S &= \frac{ 4 \lambda_m  \lambda_r  \theta  (p-p_c)   \sum_{i=1}^n q_i+ 4 A \lambda_m \mu_r^2 \sum_{i=1}^n D_i  }{  2 \lambda_m \lambda_r \rho +  \sqrt{ \triangle^S }},\\
     C^S&= \frac{p_c F_0}{\rho} +\frac{{B^S}^2 \mu_m^2}{2 \lambda_m \rho} +\frac{ 4 n B^2 \mu_r^2 - 4 B^S \mu_r^2 \sum_{i=1}^n {D_i^S}^2 + \mu_r^2 \sum_{i=1}^n D_i^S}{8  \lambda_r \rho },\\
     D_i^S&=\frac{ 4 \lambda_m \lambda_r \theta p_i q_i}{2 \lambda_m \lambda_r \rho + A^S \lambda_m \mu_r^2 + \sqrt{ \triangle^S }},\\
     H_i^S&={\frac{ D_i^S \mu_r^2 (4nB+ 2\sum_{i=1}^n {D_i^S} - 2B-D_i^S)}{4  \rho \lambda_r}} +\frac{B^S D_i^S \mu_m^2}{\rho \lambda_m}.
\end{align}}
where $\triangle^S = (4 \delta \lambda_m \lambda_r  + 2 \lambda_m \lambda_r \rho -4   \lambda_r \mu_m  p_c  \omega \theta \sum_{i=1}^n q_i )^2 - 16 \lambda_r (  \lambda_r \mu_m^2  + n \lambda_m \mu_r^2 ) ( p_c \theta \omega \sum_{i=1}^n q_i)^2$.
Substituting the optimal efforts above into the dynamic function, we can obtain the trajectory of low carbon reputation and the steady-state low carbon reputation over time as follows:

\small{
\begin{equation}\label{gs}
    \begin{aligned}
    G(t)^S = G_s^S+(G(0)-G_s^S) e^{-\delta t},
    \end{aligned}
\end{equation}

\begin{equation}\label{gss}
    \begin{aligned}
    G_s^S=\frac{\mu_m E_m^S+\sum_{i=1}^n \mu_r E_{r_i}^S}{\delta}=\frac{ 4 B^S \lambda_r \mu_m^2 + 4 n B^S\lambda_m \mu_r^2 + 2 \lambda_m \mu_r^2 \sum_{i=1}^n D_i^S  }{ \sqrt{ \triangle^S }- \lambda_m \lambda_r \rho}.
    \end{aligned}
\end{equation}}
Finally, by substituting $G_s^S$ into $V_m^S(G)$, we obtain the net profit of manufacture and retailer:
{\small
\begin{align}\label{as}
V_m^S&= A^S {G_s^D}^2 + B^S G_s^S + C^S,\\
V_{r_i}^S&= D_i^S G_s^S + H_i^S.
\end{align}}

This completes the proof.
$\blacksquare$

%
%
%
%





\end{document}